\documentclass[12pt]{iopart}
\expandafter\let\csname equation*\endcsname\relax
\expandafter\let\csname endequation*\endcsname\relax
\usepackage{graphicx}
\usepackage{amsmath}
\usepackage{mathrsfs}
\usepackage{subfigure}
\usepackage{longtable}
\usepackage{dcolumn}%
\usepackage{bm}%

\newcommand{\pDecoy}{p_{\text{Decoy}}} 
\newcommand{\etaD}{\eta_\text{D}}

\newcommand{\fQubit}{f_{\text{Q}}}

\newcommand{\epsilonQKD}{\varepsilon_{\text{QKD}}}

\newcommand{\epsilonSecr}{\varepsilon_{\text{sec}}}
\newcommand{\epsilonMAC}{\varepsilon_{\text{MAC}}}
\newcommand{\epsilonEC}{\varepsilon_{\text{EC}}}
\newcommand{\epsilonVER}{\varepsilon_{\text{VER}}}
\newcommand{\epsilonPA}{\varepsilon_{\text{PA}}}

\newcommand{\epsilonSmooth}{\varepsilon_{\text{Smooth}}}
\newcommand{\epsilonVisPE}{\varepsilon_{\text{VIS}}}
\newcommand{\fPA}{\beta_{\text{PA}}}
\newcommand{\fEC}{\beta_{\text{EC}}}
\newcommand{\fSmooth}{\beta_{\text{smooth}}}

\newcommand{\QBER}{Q}
\newcommand{\deltaQ}{\delta Q}
\newcommand{\Vis}{V}
\newcommand{\deltaV}{\delta V}
\newcommand{\tBob}{t_{\text{B}}}
\newcommand{\etaPE}{\eta_{\text{PE}}}

\newcommand{\muQ}{\mu}

\newcommand{\nSift}{n_{\text{SIFT}}}
\newcommand{\nPE}{n_{\text{PP}}}

\newcommand{\rSec}{r_{\text{sec}}}
\newcommand{\rSift}{r_{\text{sift}}}

\newcommand{\fSift}{\beta_{\text{sift}}}
\newcommand{\fEst}{\beta_{\text{est}}}
\newcommand{\fMAC}{\beta_{\text{aut}}}
\newcommand{\fSec}{f_{\text{sec}}}
\newcommand{\deltaLap}{\Delta}

\newcommand{\fGate}{f_\text{gate}}

\newcommand{\Ddata}{\text{SPD}_{\text{D}}}
\newcommand{\Dmon}{\text{SPD}_{\text{M}}}

\newcommand{\etaIM}{\eta_\text{IM}}
\newcommand{\nBit}{n_\text{bit}}

\begin{document}
%
\title{A fast and versatile QKD system with hardware key distillation and wavelength multiplexing}
\author{Nino Walenta$^1$, Andreas Burg$^3$, Dario Caselunghe$^2$, Jeremy Constantin$^3$, Nicolas Gisin$^1$, Olivier Guinnard$^1$, Raphael Houlmann$^1$, Pascal Junod$^4$, Boris Korzh$^1$, Natalia Kulesza$^2$, Matthieu Legr\'e$^2$, Charles Ci Wen Lim$^1$, Tommaso Lunghi$^1$, Laurent Monat$^2$, Christopher Portmann$^{1,6}$, Mathilde Soucarros$^2$, Patrick Trinkler$^2$, Gregory Trolliet$^5$, Fabien Vannel$^5$, Hugo Zbinden$^1$}
\address{$^1$ Group of Applied Physics-Optique, University of Geneva, Chemin de Pinchat 22, 1211 Geneva, Switzerland}
\address{$^2$ idQuantique SA, Chemin de la Marbrerie 3, 1227, Geneva, Switzerland}
\address{$^3$ Telecommunications Circuits Laboratory, EPFL, 1015 Lausanne, Switzerland}
\address{$^4$ University of Applied Sciences Western Switzerland in Yverdon-les-Bains (HEIG-VD), Route de Cheseaux 1, CH-1401 Yverdon, Switzerland}
\address{$^5$ University of Applied Sciences Western Switzerland in Geneva (hepia), Rue de la Prairie 4, CH-1202 Geneva, Switzerland}
\address{$^6$ Institute for Theoretical Physics, ETH Zurich, Gloriastrasse 35, 8093 Zurich, Switzerland}
\eads{\mailto{nino.walenta@unige.ch}, \mailto{hugo.zbinden@unige.ch}}
\date{\today}
\begin{abstract}
We present a 625~MHz clocked coherent one-way quantum key distribution (QKD) system which continuously distributes secret keys over an optical fibre link. To support high secret key rates, we implemented a fast hardware key distillation engine which allows for key distillation rates up to 4~Mbps in real time. The system employs wavelength multiplexing in order to run over only a single optical fibre and is compactly integrated in 19-inch~2U racks. We optimized the system considering a security analysis that respects finite-key-size effects, authentication costs, and system errors. Using fast gated InGaAs single photon detectors, we reliably distribute secret keys with rates up to 140~kbps and over 25~km of optical fibre, for a security parameter of $\epsilonQKD=4\cdot 10^{-9}$.
\end{abstract}
\maketitle
\section{Introduction}
Today's society relies heavily on confidential and authenticated communication. Encryption and authentication can be realized with provable information-theoretic security, derived from Shannon's theory~\cite{Shannon48}. This means that even an adversary who has unlimited computing powers can decipher an encrypted message or forge an authenticated message only with arbitrarily small probabilities. To date, the only message encryption scheme that has been proven information-theoretically secure~\cite{Shannon48} is the Vernam one-time pad cipher~\cite{Vernam26}. Secure message authentication has been demonstrated for schemes utilizing universal hash functions~\cite{Carter1979143,Wegman1981265}. The fundamental resources of these schemes are random and secret strings of bits, shared between the two distant parties commonly known as Alice and Bob. Hence, information-theoretically secure communication necessitates continuous distribution of random secret keys with provable security. Classically, the generation of two identical key streams of truly random bits at two distinct locations relies on the assumption of a secure channel or public-key cryptography. However, their security is based on certain assumptions, such as the difficulty to factorize large composite integers, or to compute discrete logarithms in certain finite groups.

A completely different approach is quantum key distribution (QKD), introduced in 1984 by Bennett and Brassard~\cite{Bennett84} (see Ref.~\cite{Gisin2002b} for a review). The idea is to send random bits encoded on non-orthogonal states of single photons. The security is based on the laws of quantum mechanics, in particular the no-cloning theorem which forbids the creation of identical copies of unknown quantum states and the fact that a measurement of an unknown quantum state inevitably disturbs it. Subsequent authenticated communication between Alice and Bob enables a measure of the information an eavesdropper potentially possesses, and hence, its reduction. Seen in this light, QKD is essentially a key expansion scheme, that is, a short initial authentication key is sufficient to generate continuously new information-theoretically secure keys~\cite{Gisin2002b}. Most importantly, the secret keys generated by QKD are universally composable, which allows one to partially reuse them for authenticating the distillation processes of subsequent QKD rounds. Remaining bits are then available for message encryption and authentication. QKD may also be used to enhance security of cryptography schemes based on computational complexity, e.g., AES (Advanced Encryption Standard) can benefit from regularly refreshed encryption keys.

Since the mid 1990's, QKD has progressed rapidly in several aspects. Starting from the early demonstration of feasibility experiments~\cite{Bennett92Exp, Muller1997}, faster and faster (with bit rates on the order of Mbps~\cite{Sasaki12, dixon10}) and long reaching systems (up to 250km~\cite{Stucki2009, Wang:12}) have been developed. However, most of the early experiments focused only on the physical layer: photon generation, manipulation, transmission and detection. Even up to today, systems which include all necessary components for secure and fast QKD are rare. Indeed, those components are numerous and need multidisciplinary competences (see Fig.~\ref{fig:myPaperOverview}). Important and often forgotten parts include random number generation, real-time error correction and privacy amplification, secure authentication and finite-key security analysis. 

In this paper, we present the results of a project~\cite{nanoTera2009} whose ambition was to implement a complete and practical fiber based QKD prototype in collaboration between six research teams in Switzerland. In particular we put the emphasis on continuous operation with a wavelength multiplexed service channel for synchronization and distillation, efficient hardware real-time distillation, finite-key security analysis, and frugal authentication. In section 1, we present the heart of any QKD prototype, the FPGA based engine controlling all the hardware as well as the complete key distillation and authentication process. This QKD engine can be adapted to many QKD protocols. In section 2, we briefly present the employed ``coherent one-way'' (COW) protocol and its specific opto-electronic realisation. Section 3 presents the experimental results and a discussion.
\section{QKD engine}
\begin{figure*}[tbp]
	\begin{minipage}{\columnwidth}
		\centering
		\includegraphics[width=\columnwidth]{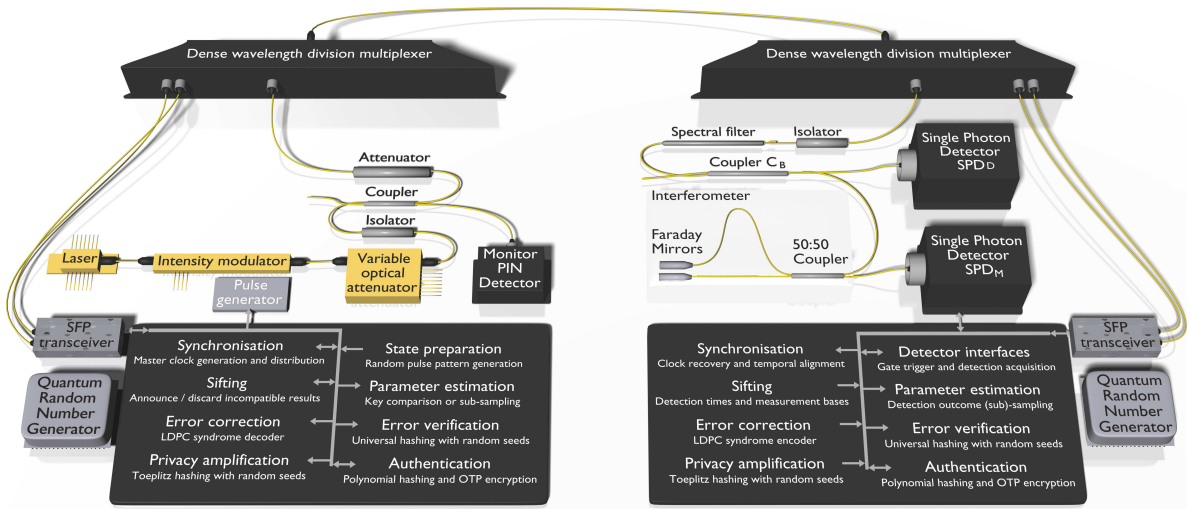}
	\end{minipage}
\caption{Schematic of our optical implementation for the coherent one-way QKD protocol and the key distillation procedures implemented in the fast FPGA hardware.}
\label{fig:COWScheme}
\end{figure*}

The QKD system described in the following was designed to have the flexibility to adapt to different QKD implementations and protocols. A schematic of our implementation is shown in Fig.~\ref{fig:COWScheme}. It is built around FPGAs (field programmable gate array, Xilinx Virtex 6) and manages the fast interfaces for the optical components, the classical communication channels, all the sub-protocols which accompany QKD as well as the distribution of the generated secret keys. The choice of the various parameters as well as all the algorithms used for key distillation and authentication processes have been carefully chosen by taking into account various trade-offs between engineering and cost constraints. Importantly, we have taken special care to analyse and optimise all tasks with respect to reducing the requirements and resources such that only one single FPGA is needed in each device. In general, compromises had to be found between the post-processing key size ($\geq 10^{5}$~bits), as required in finite-key scenarios analyzed in~\ref{sec:Methods_PREP}, and limits imposed by the hardware in terms of memory size and throughput. 
A personal computer (PC) is connected to each FPGA via PCI Express to access the configuration, status and monitoring registers. The final secret key can be transfered from the key manager to this PC and further distributed to external applications. Two communication links are established, a one-way quantum channel and a bidirectional classical service channel. All channels can be wavelength-multiplexed on a single fibre using DWDM.
In the following, we describe in more detail the functionality of each module of our QKD engine. For a more complete (and technical) description of the architecture of the code and the used algorithms, please refer to 
~\cite{RHJC2013}.

\textbf{Quantum channel interface module:} Two digital 1.25~Gbps serial interface transceivers at each FPGA (for Alice and Bob) allow synchronised interconnection with the optical hardware of the quantum channel. 
At Alice, they output up to two parallel streams of digital on-off pulses with adjustable amplitude and width, which are used to drive an electro-optical modulator for quantum state preparation. For the implementation of the COW protocol as presented later, one transceiver is needed to drive an intensity modulator. Using the second transceiver as well, one can control a dual-drive modulator and prepare all quantum states required by BB84 or the differential phase-shift (DPS) protocol, as we have shown in~\cite{Boris2013}. 
At Bob's device, both digital transceivers are used, each connected to one single photon detector $\Ddata$ and $\Dmon$, respectively. They provide the detector gate trigger if needed, and receive the detection signals from the corresponding single photon detector. Digital delays with 10~ps resolution allow temporal alignment of the detector gates with respect to the quantum signals, and temporal alignment of the detection signals with respect to Bob's FPGA clock.

\textbf{Service channel interface module:} Two optical 2.5~Gbps SFP (Small form-factor pluggable) transceivers (Finisar) on each side establish a bidirectional classical communication link between Alice and Bob.  All tasks which are needed to continuously generate secret keys or to further use these keys, share this link employing temporal multiplexing. These tasks requiring classical communication comprise in particular synchronisation, alignment, sifting, parameter estimation, error correction and verification, privacy amplification, authentication, key management, encryption, administration, and logging. Some of them strictly require authentication, some of them encryption, or even both as discussed later. The priority of each task, as well as the allocated communication bandwidth, can be adjusted individually. We employ dense wavelength-division multiplexing to transmit all classical communication channels together with the quantum channel simultaneously over a single fibre.
The FPGA system clock of Bob is synchronised and phase stabilised with some 10~ps precision with the master clock at Alice. All other necessary frequencies are derived from this clock, most importantly Alice's quantum state modulation frequency and Bob's detector gate frequency.

\textbf{Sifting and sampling module:} This module realises sifting of incompatible detections and optionally parameter estimation. Sifting comprises essentially three steps. First, since a large fraction of photons is lost in the fibre link or is not detected, Bob discloses which of the qubits he detected, without revealing the detected bit value. Second, Bob announces for each detection his randomly chosen measurement basis. Finally, Alice responds for each detection whether or not to discard it due to incompatible preparation and measurement basis. The first two sifting steps have to be performed as fast as possible in order to allow Alice to sift out undetected and incompatible bits from her memory before exceeding the available buffer size. In each sifting block, Bob encodes the detection time index of a detection relative to the index of the previous detection. Additionally, he attaches to each sifting block two control bits, which are used to indicate either the measurement basis for each detection, or empty blocks when no detection occurred during the maximum time that can be encoded in a single sifting block.

\begin{figure*}[tbp]
\begin{minipage}{\textwidth}
		\centering
		\includegraphics[width=0.48\columnwidth]{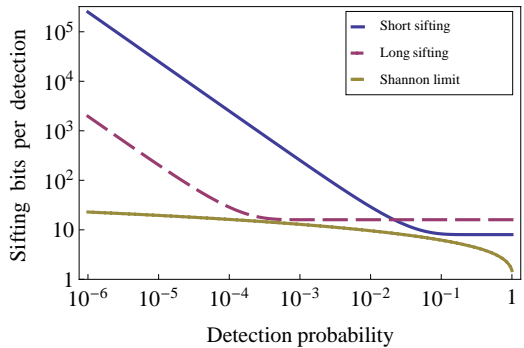}
\end{minipage}
	\caption{Number of bits per detection which have to be sent from Bob to Alice for detection times and base sifting. Blue corresponds to short sifting blocks optimised for detection probabilities $>0.021$, red uses longer sifting blocks optimised for lower detection probabilities. For comparison, the minimum amount given by the Shannon limit is shown in yellow (dashed).}
	\label{fig:SiftingBits}
\end{figure*}

The amount of bits exchanged during sifting has to be kept as small as possible, since this communication has to be authenticated at the cost of secret bits. The longer the fibre, the more bits are needed to indicate the time (number of clock cycles) passed between two succeeding detections.  We switch to 14~bits instead of 6, for detection probabilities smaller than $2\cdot10^{-2}$~per gate. As shown in Fig.~\ref{fig:SiftingBits}, our way to encode the time information is very efficient (less than twice the Shannon limit) for detections probabilities between $10^{-1}-10^{-4}$ per gate.

Some QKD protocols, e.g. COW, use only one basis to obtain the raw key. All detection outcomes in the second basis are publicly revealed in order to estimate the \textit{phase error} of the received quantum states. Bob reveals these measurement outcomes in the two control bits, too. If parameter estimation based on randomly revealing a fraction of detection outcomes is required for the quantum bit error rate (QBER) in the raw key, optionally a third control bit can be sent per detection. However, for the results we present here, we omit such sampling in favor of a more efficient solution as described below.

If double detections occur in both detectors at the same time, we only keep the result from one detector, e.g. for COW the data detector $\Ddata$. If double detections occur in both time-bins of the same qubit, we assign a randomly chosen value. A logical deadtime between $8$~ns and $10~\mu$s can be applied after detection, during which all detection are discarded to reduce impairment due to detector afterpulsing.

\begin{figure*}[tbp]
	\begin{minipage}{0.48\textwidth}
		\centering
		\includegraphics[width=\columnwidth]{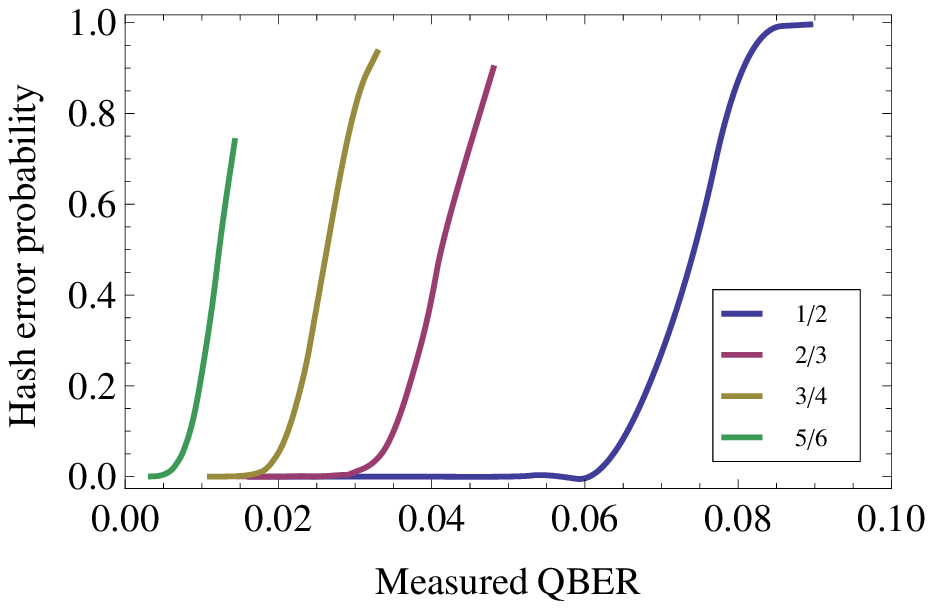}
	\end{minipage}
\vspace{0.5cm}
	\begin{minipage}{0.48\textwidth}
		\centering
		\includegraphics[width=\columnwidth]{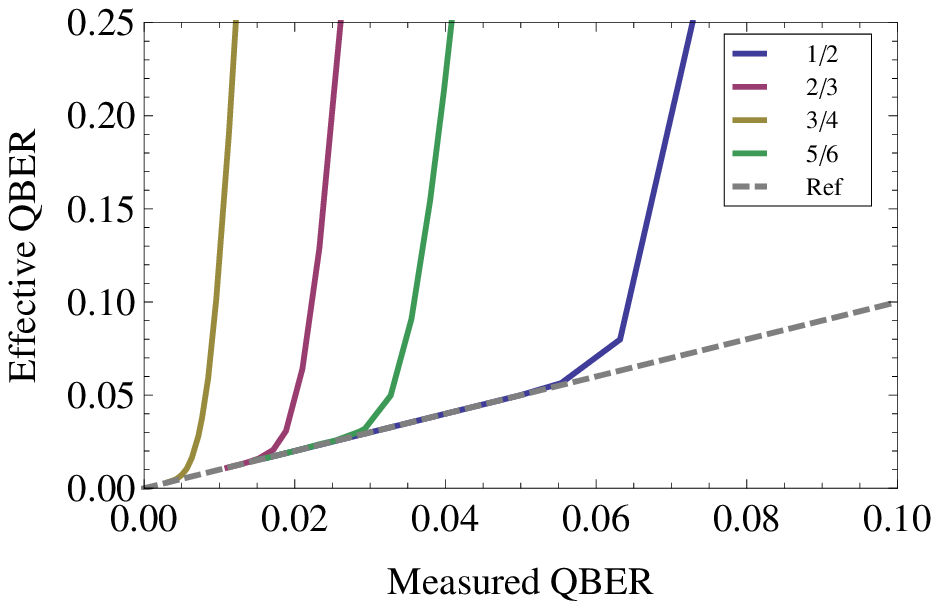}
	\end{minipage}
\caption{Left: Measurement results for different code rates showing the probability that the comparison between Alice and Bob's verification hash tags indicates at least one remaining error per 2048~bit block of error corrected keys. Right: Effective QBER under the conservative assumption that during each block with verification hash failure the eavesdropping attacks induced an error rate of~$1/2$.}
	\label{fig:ldpcPerformance}
\end{figure*}

\textbf{Error correction and verification module:}  Due to practical limitations in the preparation of the quantum states, and due to detector noise and jitter, Bob's sifted key differs from Alice's original key even in absence of eavesdropping. Therefore, a forward error correction (FEC) code is implemented in the FPGA as described in~\cite{BurgLDPC2010}, which uses the quasi-cyclic LDPC (Low-density parity-check) code defined in~\cite{IEEE802.11n}. Error correction based on LDPC codes uses syndrome encoding with the advantage that only non-iterative one-way communication is required. Moreover, it's efficiency in terms of revealed information can in principle approach the Shannon limit. 
Our FPGA implementation for LDPC performs forward error correction on blocks of 1944~bits length and provides rates up to 235~Mbps at 62.5~MHz clock frequency with ten decoding iterations. The LDPC code rate, i.e., the fraction of unpublished information, can be set to  $f_{\text{EC}}\left(\QBER\right)\in\left\{1/2, 2/3, 3/4, 5/6\right\}$ to adapt to the expected error rates. Bob calculates all syndromes for a constant expected error rate, and forwards them to Alice through an authenticated channel. Alice performs syndrome decoding and checks the parity. If an error occurred, the corresponding block is discarded. However, there is still a certain probability that uncorrected errors remain after error correction, especially for error rates larger than 6~\% (see Fig.~\ref{fig:ldpcPerformance}, left). To detect remaining errors, we implement a subsequent verification step, where Bob transmits a 48-bit hash checksum per LDPC code block to Alice. The checksums are generated using polynomial hashing~\cite{Carter1979143, Wegman1981265}, with a new random 48-bit seed for each checksum. The universal hash function is randomly chosen, and the collision probability on at least one of 512~subsequent blocks (corresponding to 995,328~bits input length for privacy amplification) is upper-bounded by~$\varepsilon_{\text{VER}}\leq 7.7\cdot 10^{-11}$. For each block, the hash, as well as the random choice of hash function, are sent to Alice. If a checksum mismatch occurs, the associated block is discarded. Fig.~\ref{fig:ldpcPerformance} (left) shows for all implemented code rates the probability that a verification fails as a function of the measured raw QBER.

\textbf{Bit error estimation module:} In every QKD protocol the amount of errors of the received quantum states has to be estimated in  order to determine an upper bound on the fraction of information which could have leaked to an eavesdropper. The standard procedure consists in random sampling of a subset of the sifted key, comparing the bit values over an authenticated channel, and calculating the error rate in each basis. While straightforward, this method reduces the final secret key rate as all revealed outcomes have to be discarded. Most importantly, it has a substantial impact for a finite-key analysis, since a small sample gives only an imprecise estimate on the true error rate in the remaining, unrevealed detections.

To overcome these impairments, we perform parameter estimation exploiting our knowledge about the correctness of the key after verification. Once we obtain 512~blocks of 1944~error corrected and verified bits, Alice compares them with her original random bit sequence~\cite{Pacher2012}. By counting the total number of mismatches, an exact number for the true bit errors is obtained. Additionally, we take into account blocks which were dropped due to checksum mismatches during error verification. We conservatively assume for each block with verification hash failure a maximum error rate of $1/2$ induced by eavesdropping attacks. In Fig.~\ref{fig:ldpcPerformance} (right) we show the resulting, effective QBER for different code rates as a function of the measured QBER. The failure probability for parameter estimation is then equal to the failure probability of error verification, i.e., $\varepsilon_{\text{PE}}\leq 7.7\cdot 10^{-11}$.

\textbf{Privacy amplification module:} Our FPGA implementation of privacy amplification uses Toeplitz hashing~\cite{Carter1979143,Wegman1981265}, a construction for families of universal hash functions, in combination with LFSR hashing as proposed in~\cite{Mansour1993121, Krawczyk_1994}. This approach is very efficient in terms of communication bandwidth needed to convey the chosen hash function, and allows parallelised computation and efficient, scalable implementation on the FPGA hardware.

The privacy amplification compression is the ratio between the length of the output and input keys, i.e., the ratio between the number of rows and columns of the Toeplitz matrix. In order to obtain high secret key rates based on finite-key analysis, we choose a fixed input length of 995,328~bits. As a consequence of this large block size, the size of the resulting matrix is such that it has to be stored in an SDRAM outside the FPGA. Our hardware implementation for privacy amplification has shown to treat up to 48~Mbps input rate. Changing the output block length, the compression ratio can be adjusted over the full range between $0-100~\%$ in steps of 0.05~\%. We optimise and fix the compression ratio once in advance for a given scenario. Then, we verify for each key that the parameter estimates are indeed within the limits which guarantee security with the chosen compression ratio.

\textbf{Authentication module:} The classical communication channel is authenticated in order to prevent an eavesdropper from forging messages, which would open the door for man-in-the-middle attacks. For information-theoretically secure authentication, we use a combination~\cite{Stinson1994} of $\varepsilon_{\text{AUT}}$-almost strongly universal hash functions in combination with a strongly-universal family of hash functions named polynomial hashing~\cite{Carter1979143, Wegman1981265}, which is very efficient with respect to consumed secret bits as well as required operations. Bob randomly and secretly selects a hash function from this family to calculate a hash tag for each transmitted message, and sends the hash tags together with the messages to Alice. To verify that the transmission has not been forged, Alice has to know which hash functions Bob has chosen to be able to verify the hash tags for the received messages. Only when her calculated and the received tag for a message match, then it is considered valid. We send a new 127-bit authentication tag for every $2^{20}$~bits of classical communication to obtain a collision probability of $\varepsilon_{\text{AUT}}\leq 10^{-33}$. This approach would require 383~secret bits to select a new hash function for every tag. However, recently it has been shown that the same hash function can be reused for multiple authentication rounds if the tags attached to the messages \cite{Portmann2012} are one-time pad encrypted. This authentication scheme is proven $\varepsilon$-universal-composable-secure even if $\varepsilon$-almost strongly universal$_{2}$ hash functions are used and provides a bound for its information leakage. This strategy reduces the secret key consumption to one third, since only 127~bit secret keys are needed to encrypt each tag instead of 383~secret bits to select a new hash function.

\textbf{Random number generation module:} Random numbers are extensively needed during preparation for selecting the quantum states, as well as during key distillation, e.g., to generate the privacy amplification matrices. These random bits must be provided by true quantum random number generators, ideally quantum random number generators (QRNGs) where up to 2~GHz output rates have been demonstrated~\cite{symul2011} to date. However for the time being, we use a commercial QRNG~\cite{Quantis} (certified by Swiss Federal Office of Metrology). Since its bit rate of 4~Mbps is by far not sufficient, we implement the NIST SP800-90 recommended AES-CTR cryptographically secure pseudo-random number generator that uses seeds of 256~bits provided by the QRNG to generate up to 1.1~Gbps random bits. We note that due to AES, the random number expansion protocol is the only part of the entire system for which we can't provide an information-theoretic security statement.

\textbf{Key manager:} A fraction of the privacy amplified, secret keys is transfered by the key manager to the authentication module. Once their authenticity has been verified, the key manager distributes the remaining keys to an internal OTP encryption application, or via a PCI Express link to a PC and further to external consumers, e.g network encryptors.
\section{COW protocol and implementation}
The presented QKD system provides the flexibility to drive different QKD protocols ~\cite{Boris2013}. In the following, we present the implementation of the coherent one-way (COW) protocol ~\cite{stucki-2005-87}.

\textbf{The COW protocol} belongs to the class of distributed phase reference protocols and seeks to enable long fibre distance QKD while maintaining a simple and convenient setup. The advantages of the COW protocol are that it allows implementation of a completely passive receiver, without any active element for base choice, requiring only two single photon detectors. Its implementation is robust against birefringence fluctuations, fibre transmission losses and photon number splitting attacks. A schematic of the setup is sketched in Fig.~\ref{fig:COWScheme}. 

Following the COW protocol, Alice encodes each bit value by the choice of sending a weak coherent pulse in one out of two possible time-bins, while the other time-bin contains the vacuum state. Formally, these quantum states can be written as $\left|\beta_{\text{0}}\right\rangle_{\text{n}}=\left|\alpha\right\rangle_{\text{2n}}\left|\text{vac}\right\rangle_{\text{2n-1}}$ and $\left|\beta_{\text{1}}\right\rangle_{\text{n}}=\left|\text{vac}\right\rangle_{\text{2n}}\left|\alpha\right\rangle_{\text{2n-1}}$, where $\alpha$ is the complex coherent state amplitude with an average photon number per time bin $\mu=\left|\alpha\right|^{2}<1$, and $n$ labels the qubit index.  These states can be discriminated  optimally by a simple time-of-arrival measurement. In addition, a third state called decoy sequence with both time-bins containing weak coherent pulses is randomly prepared, i.e. $\left|\beta_{\text{d}}\right\rangle_{\text{n}}=\left|\alpha\right\rangle_{\text{2n}}\left|\alpha\right\rangle_{\text{2n-1}}$. 

As for distributed-phase-reference QKD, the integrity of the quantum channel is monitored using an imbalanced interferometer. It measures the coherence between pulses in two successive, non-empty time-bins, either within a bit when a decoy sequence was prepared, or across bit separation whenever corresponding sequences are prepared. Latter measurement across bit separation renders  photon number splitting attacks on individual states less powerful as the adversary reduces the interference visibility if trying to discriminate individual states. As a consequence, the optimal average number of photons which can be sent per qubit becomes independent of the fibre transmission, but dependent on QBER and visibility. Security against zero error attacks and restricted collective attacks was proven, including imperfections of the state preparation~\cite{Branciard2008}. Note, that a general security proof was obtained for a modified COW protocol~\cite{Moroder12}, which, however, involves more intricate hardware.

\textbf{Alice's optical QKD module:} The coherent light source is  a continuous-wave distributed feedback laser diode (Agilecom) with a sufficiently long coherence time of~$>300$~ns. It is compatible with 100~GHz DWDM telecom standard, and its central wavelength regulated by a thermo-electric controller (TEC) to $\lambda = 1551.72$~nm (ITU channel~$32$)~\cite{ITUCGrid}.

An integrated LiNbO$_{3}$ intensity modulator (IM, Photline MX-LN~20) prepares the COW states. It tailors the continuous optical signal in a coherent train of short pulses, according to the states selected by the random number generator. The corresponding digital on-off signals are provided through the high-speed serial interfaces of the FPGA, reshaped to clean pulses of $50-400$~ps duration, and amplified to appropriate voltage levels for the IM input. The bias voltage is adjusted to maximise the optical pulse extinction ratio. Indeed, the extinction ratio of the IM limits the minimum quantum bit error rate since spurious light in a supposedly empty time bin causes erroneous detections. Therefore, we use the QBER as feedback to re-adjust the IM bias voltage continuously. More than $25$~dB extinction is achieved for $130$~ps long pulses at a frequency of 625~MHz, limiting the expected QBER to 0.3~\%. Decoy sequences are prepared with a probability of 15.5~\%, close to the optimum, which allows for a sifted key rate as high as 73~\% of the raw key rate.

A MEMS based variable optical attenuator (Sercalo) attenuates the quantum signal down to the optimal photon level at Alice's output. Its value is optimised with respect to the QBER, visibility and other parameters as discussed later. The optical isolator prevents Trojan horse attacks (based on sending bright light from the outside). A 90:10 imbalanced fibre coupler and tap monitor diode allow continuous monitoring of Alice's output power and providing feedback to the variable optical attenuator to adjust the average number of photons per bit. Moreover, an unexpected increase of power in the monitor diode would indicate malfunction or a Trojan horse attack. Finally, a fixed, calibrated optical attenuator just before Alice's output reduces the average photon number per pulse to the optimal value.

\textbf{Bob's optical QKD module:} At Bob's quantum channel input, an optical isolator prevents information leakage due to detector backfiring or back-reflection of potential Trojan horse attacks. A 45~pm spectral fibre Bragg grating (FBG, aos) filter with $1.4$~dB insertion loss and $14$~dB isolation reduces incoming Raman noise. Subsequently, a fibre coupler realises the passive, random base choice and splits the quantum signals towards data and monitoring line. Its splitting ratio of 80:20 is close to optimal for most experimental settings used in the following.

Two single photon detectors are installed: $\Ddata$ measures the photon arrival time in the data line to obtain the raw key, $\Dmon$ detects the output of the imbalanced interferometer (IF) in the monitoring line.
For the results presented in sec.~\ref{sec:experimentalResults}, $\Ddata$ is a sine gated InGaAs avalanche photo diode (APD) with a frequency of 1.25~GHz as described in \cite{walentaSine}. Its gate width (FWHM) is 130~ps which proves to be a good tradeoff between sufficiently low afterpulsing while maintaining a good detection efficiency. The efficiency is varied in the range of 6-10~\%, maximising the final secret key rate. For the considered fibre distances, the dark counts are no limiting factor and the highest key rate was indeed obtained at room temperature  ($20^{\circ}$C). At this temperature, the dark count probability is about $10^{-6}$ per gate at 10~\% efficiency.

As the monitoring detection rate is much smaller, $\Dmon$ is a free-running negative feedback InGaAs APD~\cite{Lunghi2012}. Applying 20~$\mu$s deadtime, its dark count rate was typically 800~Hz at 20~\% detection efficiency. Importantly, its timing jitter is only 200~ps (FWHM), sufficiently low to discriminate time-bins at 1.25~GHz. The gate times for both detectors are derived from the clock signal distributed over the service channel, and are digitally delayed to compensate for any temporal delay between quantum and service channel.

The Michelson type IF as sketched in Fig.~\ref{fig:COWScheme} is made up of a fibre coupler with two Faraday mirrors terminating the two arms. The arms are cut such that its length difference corresponds to half of the separation between consecutive time bins. The measured free-spectral range of $1.247$~GHz matches very well the target frequency of $1.25$~GHz. The IF has $1.3$~dB insertion loss, and a maximum visibility $>0.998$. It is thermally well isolated and actively temperature stabilised. The relative phase, however, is adjusted by tuning Alice's laser wavelength such that two succeeding pulses interfere destructively and don't generate detector clicks. In contrary, non-interfering pulse sequences are distributed randomly between the two output ports of the IF. Note, the second output port can be monitored via an additional circulator at the cost of increased insertion loss and the need of a third detector. This would slightly increase the secret key fraction, as Eve's information could be estimated more precisely. 

\begin{figure*}[tbp]
		\centering
		\includegraphics[width=\textwidth]{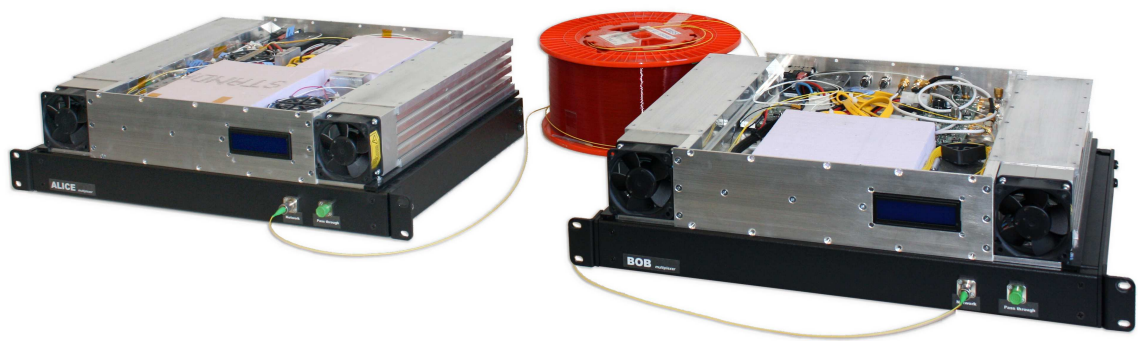}
	\caption{Photo of the opened QKD devices. Each system is compatible with 19-inch~2U industrial cases and houses all the electronics, optics and interfaces to distribute quantum keys, use the QKD keys for Ethernet authentication and one-time pad encryption, and to additionally supply them to external consumers devices. In consideration of security aspects, their interior is completely mechanically encapsulated, while thermal stabilisation is provided by two external fans. Using external 19-inch~1U DWDM modules (bottom), both devices were connected by only one single telecom fibre and have demonstrated stable QKD functionality with a security guarantee of $\epsilonQKD=4\cdot 10^{-9}$ over more than 25~km distance.}
	\label{fig:myPaperOverview}
\end{figure*}

\textbf{Mechanical housing and DWDM modules:} Each QKD device is integrated in a 19-inch~2U housing as shown in Fig.~\ref{fig:myPaperOverview}. It provides a power input, a single mode fibre connector (APC) for the quantum channel, a PCI-Express link to the control PC, and two SFP slots for the service channel and an optional external encryptor. Importantly, despite these connectors the mechanical housing is perfectly encapsulated from the environment to prevent any other physical attack point than through the optical fibre. In particular, the arrangement of all components has been carefully chosen to maintain an efficient heat release and to guarantee maximum stability, although the cooling air flows only outside around the device without entering it.

During all key exchanges presented here, we used one single optical fibre and dense wavelength-division multiplexing (DWDM) for quantum and all classical channels. We implemented external DWDM modules for Alice and Bob in separate 19-inch~1U cases, comprising a 100~GHz multiplexer (OptiWorks) and a variable optical attenuator (OptoLink) to minimise the power of the transmitted classical channels. The multiplexers have an isolation of $80$~dB and an insertion loss of $1.8$~dB.
\section{Experimental results}\label{sec:experimentalResults}
We tested the system over fibre lengths between $1-50$~km using rapid sine gated single photon detectors~\cite{walentaSine} as well as free-running single photon detectors (id220, IDQ). All classical and quantum communication channels were multiplexed onto a common fibre. Using different configurations of the distillation engine we optimised the key rates for a security parameter of $4\cdot 10^{-9}$, while respecting a security analysis for finite-key-size effects, authentication costs, and system errors.

\begin{figure*}[tbp]
	\begin{minipage}{0.48\textwidth}
		\centering
		\includegraphics[width=\columnwidth]{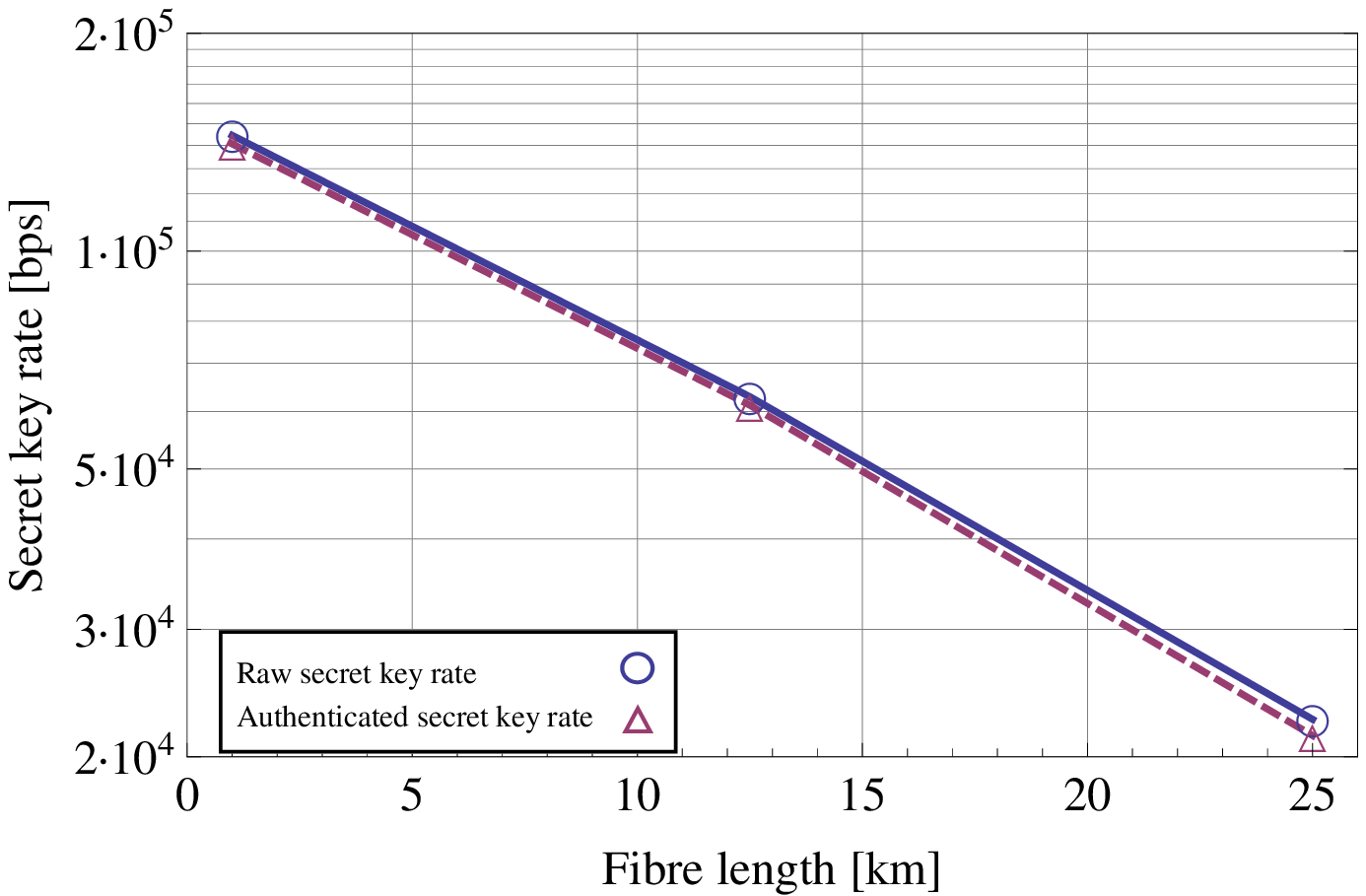}
	\end{minipage}
\vspace{0.5cm}
	\begin{minipage}{0.48\textwidth}
		\centering
		\includegraphics[width=\columnwidth]{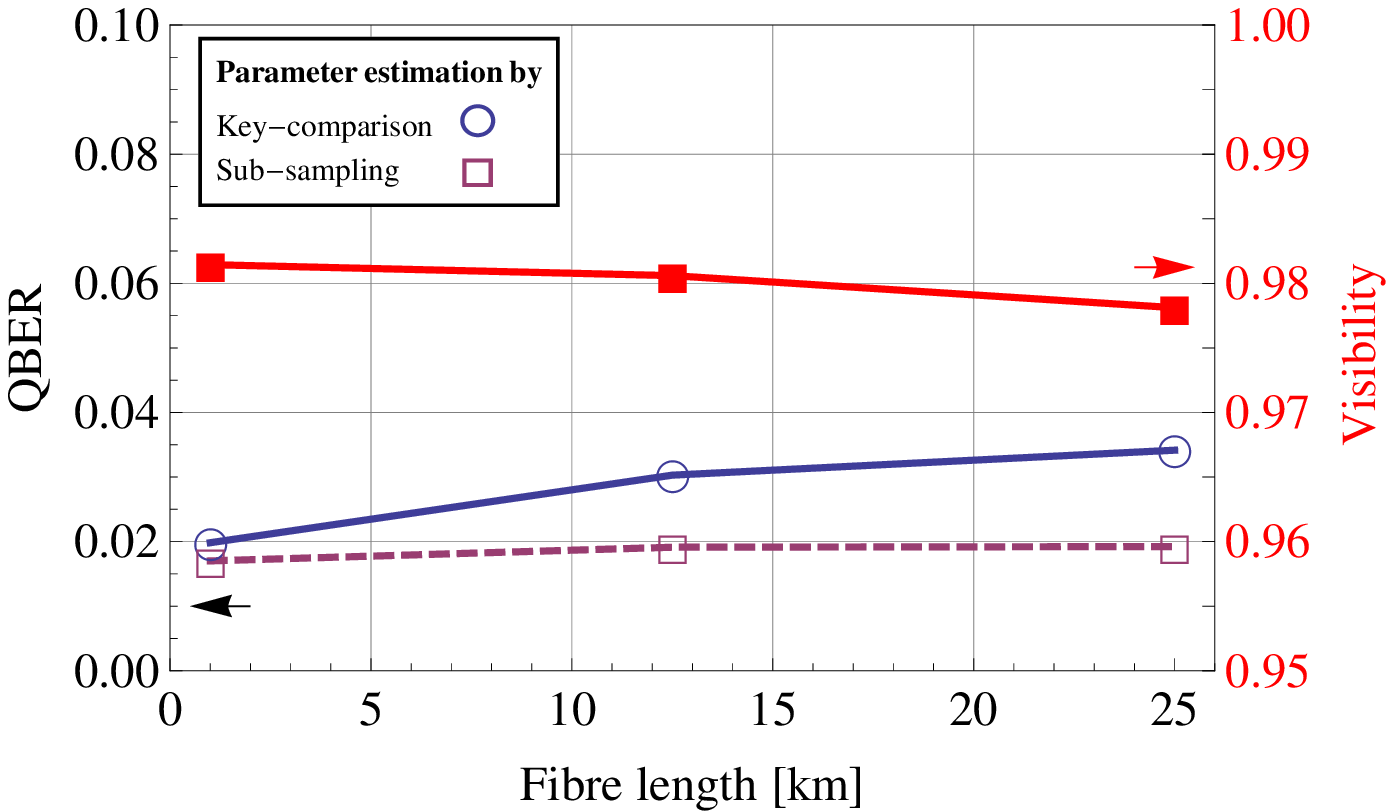}
	\end{minipage}
	\caption{Left: Secret key rates after privacy amplification (blue circles) and authenticated secret key rate (purple triangles) which accounts for secret key consumption for authenticating the classical communication channel. We considered a security analysis that respects finite-key-size effects, authentication costs, and system errors with a security parameter of $\epsilonQKD=4\cdot10^{-9}$. Right: QBER and raw visibility results before removing dark counts.}
	\label{fig:mainSKRs}
\end{figure*}

For the measurements which we discuss in the following, we obtained the highest secret key rate using an LDPC error correction code rate of $3/4$, parameter estimation based on key comparison, and longer sifting blocks to encode the detection times in 14~bits. The secret key rate which is provided by the FPGA distillation engine after privacy amplification, is shown in Fig.~\ref{fig:mainSKRs} (left, circle). Multiplexing quantum and classical channels over a single 1~km fibre, secret keys were distributed at a rate of 144.5~kbps. Over a single 25~km long fibre, we obtained after privacy amplification a secret key rate of 22.5~kbps. The useful rate of secret bits available for applications, e.g., internal one-time-pad encryption or external encryptors is shown as red triangle in Fig.~\ref{fig:mainSKRs} and accounts for secret bit consumption to encode the authentication tags.

\begin{figure*}[tbp]
	\begin{minipage}{0.48\textwidth}
		\centering
		\includegraphics[width=\columnwidth]{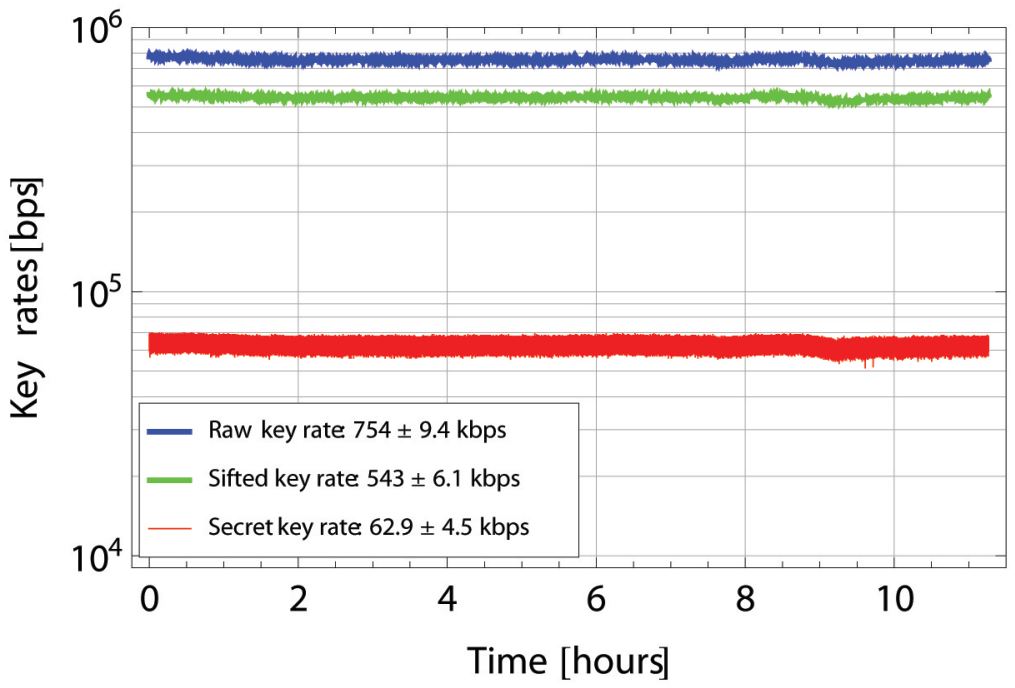}
	\end{minipage}
\vspace{0.5cm}
	\begin{minipage}{0.48\textwidth}
		\centering
		\includegraphics[width=\columnwidth]{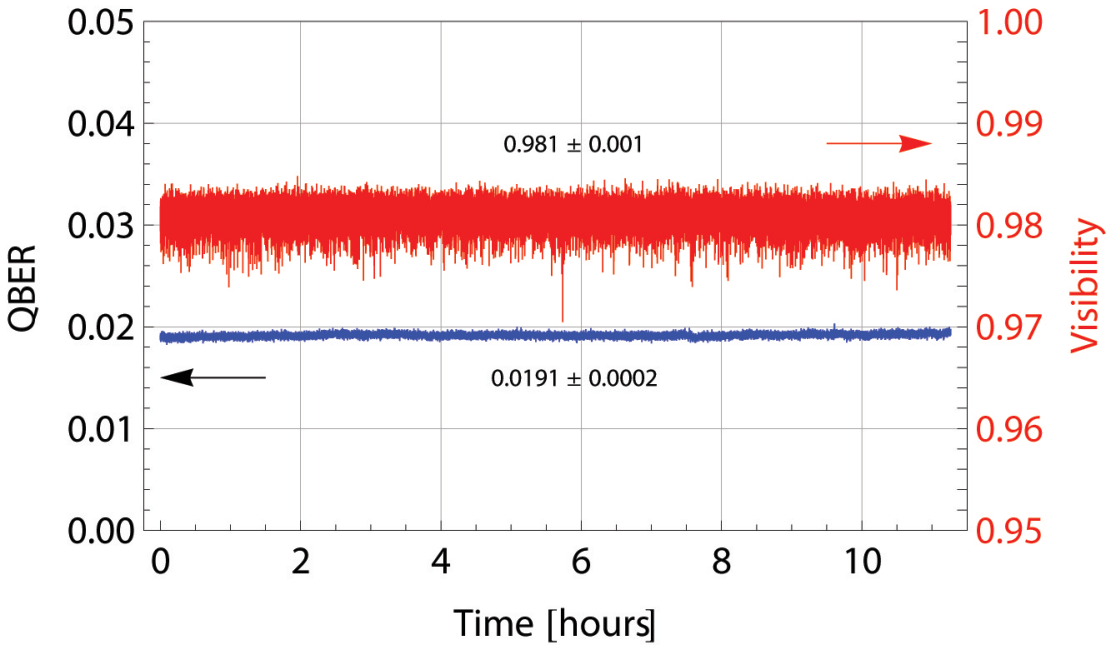}
	\end{minipage}
	\caption{Key rates (left), QBER and visibility (right) demonstrating the stability of an autonomous QKD run for a period of more than 11~hours. Alice's and Bob's devices were connected by a single 12.5~km fibre. The secret key rate (left, red) accounts for finite-key effects, the authenticated key rate (left, purple) for the consumption of secret keys to encrypt the authentication tags.}
	\label{fig:mainLongs}
\end{figure*}

\textbf{Parameter optimisation:} For each setting we optimised several parameters to maximise the final authenticated secret key rate. These are summarised in Table~\ref{tab:summary}. For longer fibres, the average photon number was increased and the detection efficiency decreased in order to compensate for increasing impairment due to DWDM noise (Raman scattering and crosstalk) and dark counts. Such, the quantum bit error rate was maintained close to the maximum QBER which could be efficiently corrected with the chosen LDPC code rate (see Fig.~\ref{fig:ldpcPerformance}). For the different fibre lengths we obtained a QBER (before subtracting dark counts) as shown in Fig.~\ref{fig:mainSKRs} (right). The QBER  increases for longer fibres and is considerably larger than the error rate which we estimated using sub-sampling instead. This additional contribution stems from blocks of error corrected bits, which haven't passed the subsequent hash tag verification. For these blocks we conservatively attribute a-priori an error rate of $1/2$ to the eavesdropper. Thus, with a verification failure probability of $3.1$~\% for a 25~km fibre, the QBER which we take into account increases above 3.4~\%. Nevertheless, we verified that in the presented configurations the final secret key rate was still higher compared to configurations with parameter estimation based on sub-sampling, since the impairment due to verification failures is overcompensated by the advantage that no bits have to be revelead and discarded. Similarly, we found that smaller error correction code rates did not result in higher key rates.

The raw visibility (before subtracting dark counts) in Fig.~\ref{fig:mainSKRs} (right, red) remains almost constant for all fibre lengths. It slightly drops below 97~\% for long fibres due to increasing impairment stemming from DWDM noise and dark count detections.  Mainly determined by the visibility and photon number, and with slight dependence on the QBER, we applied privacy amplification with a compression factor of 11.5~\% for a fibre of 1~km length, which dropped to 6.5~\% for 25~km.
\begin{table*}[tbp]
\begin{center}
\footnotesize
{\renewcommand{\arraystretch}{1}
\begin{tabular}{|l|c|c|c|}
\hline
Fibre length [km] & 1~km & 12.5 km & 25 km\\
\hline
Pulse amplitude $\mu$ & 0.089 & 0.084 & 0.105\\
Detection efficiency [\%] & 9.6 & 7.3 & 6.9\\
Compression factor [\%] & 11.5 & 12.0 & 6.5\\
LDPC code rate & 3/4 & 3/4 & 3/4 \\
\hline
QBER [\%] (raw/verified) & $1.70\pm 0.01$~/~$1.98$ & $1.87\pm0.02$~/~$3.03$ & $1.91\pm 0.03$~/~$3.42$\\
~~Dark count contribution & $0.41$ & $0.76$ & $0.85$\\
~~DWDM noise contribution & $0.05$ & $0.11$ & $0.19$\\
Raw visibility [\%] & $98.14 \pm 0.14$ & $98.06 \pm 0.13$ & $97.81\pm 0.13$\\
Sifted key rate [bps] & $(1.26\pm 0.006)\cdot10^{6}$ & $(5.38 \pm 0.032)\cdot10^{5}$ & $(3.59\pm 0.042)\cdot10^{5}$\\
Secret key rate [bps] & $1.45\cdot10^{5}$ & $6.29\cdot10^{4}$ & $2.25\cdot10^{4}$\\
Authenticated key rate [bps] & $1.41\cdot10^{5}$ & $6.12\cdot10^{4}$ & $2.14\cdot10^{4}$\\
\hline
\end{tabular}}
\caption{Parameters and measurement results summarizing the performance of the QKD prototype for information theoretic secure key distribution with an security parameter of $4\cdot10^{-9}$.}
\label{tab:summary}
\end{center}
\end{table*}
\textbf{Stability:} In Fig.~\ref{fig:mainLongs} we show the stability in terms of key rates, QBER and visibility for an autonomous QKD run over a period of more than 11~hours using a single 12.5~km DWDM fibre link. The results clearly reflect the good stability of all system components including synchronisation and alignment, Alice's state preparation, Bob's interferometer and single photon detectors, and the whole distillation engine. The average raw quantum bit error rate as measured by comparing Alice's error corrected key with her original key was 1.91~\% over the whole measurement period (Fig.~\ref{fig:mainLongs}, right).  The raw visibility before subtracting dark counts had an average of 98.1~\%, and was constantly above 97.0~\%. Considering finite-key security with $\epsilonQKD=4\cdot 10^{-9}$, we applied a compression factor of 0.12, and accounting for the fraction of blocks which were discarded due to verification failures, the resulting secret key rate was $62.9$~kbps.

During two live presentations at conferences, we have demonstrated the robustness, stability and reliability of our QKD system~\cite{Nanotera2013,QCrypt2013}. Over periods of 2 and 5 days, the system ran continuously and provided more than 30~times per second new secret 128-bit keys to network encryptors, which used the keys for AES encryption of user data and video streams.

\begin{figure*}[tbp]
	\begin{minipage}{0.48\textwidth}
		\centering
		\includegraphics[width=\columnwidth]{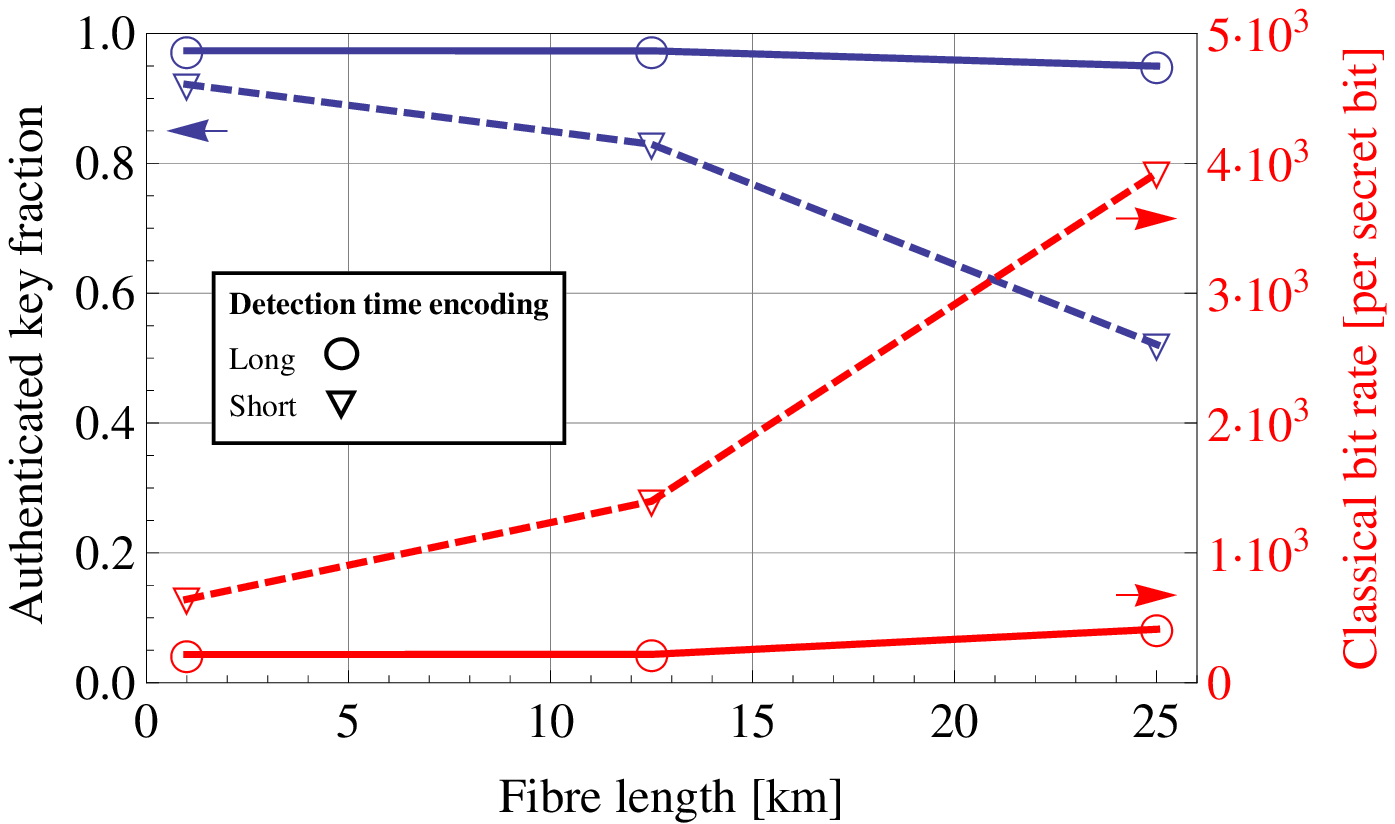}
	\end{minipage}
\vspace{0.5cm}
	\begin{minipage}{0.48\textwidth}
		\centering
		\vspace{0.5cm}
		\includegraphics[width=0.9\columnwidth]{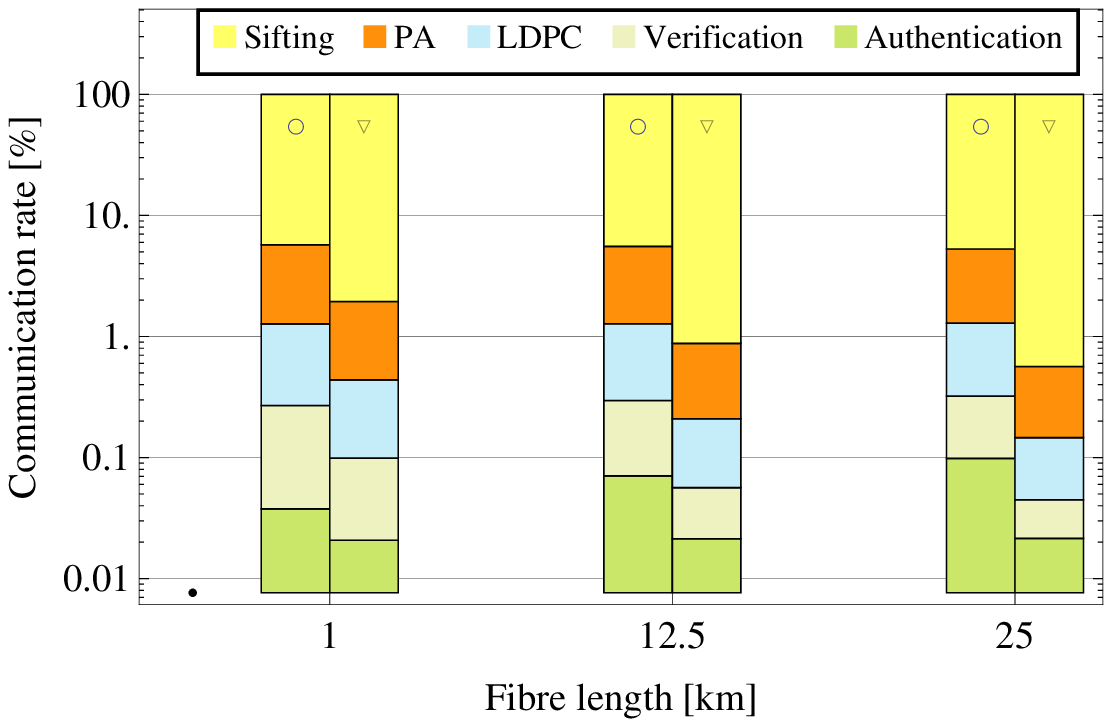}
	\end{minipage}
\caption{Amount of classical information accompanying QKD. Left: Total communication rates per secret bit and fraction of secret bits remaining after authenticating the classical communication channels. At least 2.7~\% of secret bits are consumed for authentication, i.e., to encrypt the authentication tags of 127~bits per $10^{6}$~bits of classical communication. Right: Communication rates broken down by individual sub-protocols for the considered fibre lengths. The rates are dominated by the amount of sifting information sent from Bob to Alice which adds up to $94-99$~\%, depending on the specific configuration.}
	\label{fig:ProtoAuthRates}
\end{figure*}

\textbf{Authentication costs:} The secret key rates usually presented are the key rates after privacy amplification, i.e., they do not account for secret bit consumption to encode the authentication tags. Therefore, Fig.~\ref{fig:ProtoAuthRates} shows the amount of classical communication accompanying key distillation as well as the fraction of secret bits which are consumed to encrypt authentication tags of 127~bit per $10^{6}$~bits of classical communication. The left side of Fig.~\ref{fig:ProtoAuthRates} shows the amount of classical information which has to be communicated normalised per secret bit, as well as in terms of authenticated fraction of secret bits which is left after authentication. It reveals, that for all considered fibre lengths, the least fraction of secret bits consumed for authentication is obtained if we use long sifting blocks and parameter estimation based on key comparison (circles). For a fibre of 1~km length, 217~classical bits have to be communicated per secret bit. Correspondingly, a fraction of 2.7~\% of secret bits is needed for authenticating this communication, i.e. the authenticated key rate amounts to 97.3~\%. It increases up to 412~bits of classical communication per secret bit for a 25~km fibre, where 5.0~\% of secret bits are needed for authentication, corresponding to a authenticated key rate of 95.0~\%. Much more classical information has to be sent and authenticated, if short sifting blocks with only 6~bits instead of 14~bits are used to encode the detection times, and nearly 20~\% of all secret bits are consumed for authentication (triangles in Fig.~\ref{fig:ProtoAuthRates}).

The origin of the different authentication losses is illustrated in Fig.~\ref{fig:ProtoAuthRates} (right), where we compare the communication rates broken down by each individual sub-protocol. With more than 94~\% the largest amount of information is sent for sifting. More than one order of magnitude less, up to 4.5~\%, for communicating the randomly chosen Toeplitz matrices for privacy amplification. At most $1.2$~\% of all classical communication is attributed to error correction including communication of the verification hash function and value, and less than 0.1~\% for authentication. Using shorter sifting blocks (triangles in Fig.~\ref{fig:ProtoAuthRates}), the relative amount of sifting information becomes even larger, giving raise to larger authentication loss. However, we expect that the shorter blocks used to encode the detection times become advantageous as soon as higher detection rates are obtained. This would be the case when detectors with higher detection efficiency are used, e.g., superconducting single photon detectors, or two fibre links instead of one, which would eliminate optical losses in multiplexers and spectral filters. When we used parameter estimation based on sub-sampling instead of key comparison, the amount of classical communication was 12.6~\% larger for all fibre lengths, corresponding to the fraction of bits which were revealed and discarded.
\section{Conclusions and outlook}
To conclude, we have presented a fully integrated versatile QKD platform that comprises of a hardware key distillation engine, dense wavelength-division multiplexing of quantum and all classical communication channels, and fast sine gating detectors. We demonstrated its stable performance for the coherent one-way protocol, and rigorously took into account all aspects which guarantee security in finite key scenarios with a security parameter of $4\cdot 10^{-9}$. Our QKD platform has the flexibility to not only support the coherent one-way protocol, but additionally provides all the means to run the differential phase-shift QKD protocol, as well as phase-time qubit BB84. The system is compactly mounted in standard industrial 19-inch~2U housings.

All results were obtained using a 1-fibre DWDM configuration with all quantum and classical communication channels multiplexed in one common fibre, and taking into account finite key security for a block size of $10^{6}$~bits. However, we want to stress that depending on the specific usage scenario and security requirements, the maximum secret key rate as well as the maximum fibre length can easily be increased. As an example, we performed the same set of measurements while neglecting finite-key effects, and obtained after authentication an asymptotic key rate of 293~kbps and 1.3~kbps for a fibre length of 1~km and 50~km, respectively. A further increase by more than a factor of two in both, key rate and distance, can be expected if instead of multiplexing all channels over one single fibre, two fibres are available, one dark fibre for the quantum channel and a second fibre for the classical communication channels.
\section{Acknowledgments}
We gratefully acknowledge the valuable discussions with Renato Renner, Marcos Curty and Christoph Pacher. Furthermore, we thank Herv\'e Gouraud from Photline for his kind support. This research project was financially supported by the Swiss Nano-Tera project QCRYPT and the National Center of Competence in Research QSIT.
%
\appendix
\section{COW finite-key rates}\label{sec:Methods_PREP}
We consider a coherent one-way transmitter at Alice as depicted in Fig.~\ref{fig:COWScheme} which prepares time-bin qubits with a frequency $\fQubit$. In general, the prepared quantum state after a time $t_{\mathcal{N}}=\mathcal{N}/\fQubit$ can be written in the form of a product state
\begin{align}
	&\left|\Psi\right\rangle_{\mathcal{N}} = \otimes^{\mathcal{N}}_{n=1} \left|\psi\left(b_{n}, v_{n}\right)\right\rangle_{n}\\
	&\left|\psi\left(b_{n}, v_{n}\right)\right\rangle_{n} = \otimes^{\nBit-1}_{i=0}\left|\alpha\left(b_{n}, v_{n}, i\right)\right\rangle_{n\cdot\nBit-i}
	\label{eq:globalCOWstate}
\end{align}
of coherent quantum states $\left|\alpha\right\rangle_{\mathcal{\tau}}$. Their complex amplitudes $\alpha$ in temporal mode $\tau$ depends on Alice's random choice of basis $b_{n}\in\left\{0,1\right\}$ and bit value $v_{n}\in\left\{0,1\right\}$. We have introduced a parameter $\nBit=\fGate/\fQubit$ which accounts for the implementations where $\nBit$ successive temporal modes are used to distinguish the states. It is $\nBit = 2$ for COW and BB84 phase-time qubits, while for DPS $\nBit = 1$. Whenever Alice chooses $b_{n}=0$, she prepares a quantum state corresponding to a bit value
\begin{align}
	\left|\psi\left(0,0\right)\right\rangle_{n} =& \left|\sqrt{\frac{\mu}{(1+\etaIM)}}\right\rangle_{2n} \otimes\left|\sqrt{\frac{\etaIM\cdot\mu}{(1+\etaIM)}}\right\rangle_{2n-1}\nonumber\\
	\left|\psi\left(0,1\right)\right\rangle_{n} =& \left|\sqrt{\frac{\etaIM\cdot\mu}{(1+\etaIM)}}\right\rangle_{2n}\otimes\left|\sqrt{\frac{\mu}{(1+\etaIM)}}\right\rangle_{2n-1}
	\label{eq:cowQBitStates}
\end{align}
Here, $\mu=\left|\alpha^{2}\right|$ is the mean value of the Poissonian distributed number of photons per coherent state, and $0\leq\etaIM\leq 1$ accounts for a limited extinction ratio of the intensity modulator. In the ideal case it is $\etaIM=0$, and eq.~(\ref{eq:cowQBitStates}) becomes  $\left|\sqrt{\mu}\right\rangle\otimes\left|0\right\rangle$ and $\left|0\right\rangle\otimes\left|\sqrt{\mu}\right\rangle$. Whenever Alice chooses $b_{n}=1$ with probability $\pDecoy$ a decoy sequence, irrespective of the bit value she prepares
\begin{align}
	\left|\psi\left(1,0\right)\right\rangle_{n} = \left|\psi\left(1,1\right)\right\rangle_{n} = \left|\sqrt{\mu}\right\rangle_{2n}\otimes\left|\sqrt{\mu}\right\rangle_{2n-1}
\end{align}

The goal of Alice and Bob is to maximize the COW secret key rate (per prepared state) $\rSec$ which can be distilled from the transmitted and detected states 
\begin{align}
	\rSec =& r_{\text{det}}\cdot\beta_{\text{sift}}\cdot\fEst\cdot\fSec\cdot \beta_{\text{auth}}\\
	=&\rSift\cdot\left(1-\eta_{\text{PE}}\right)\cdot\fSec\cdot \left(1-\eta_{\text{MAC}}\right)
	\label{eq:COWSecretRate}
\end{align}
where $r_{\text{det}}$ is the detection rate (per prepared bit) in Bob's detector $\Ddata$. Further, $\fSift, \fEst, \fSec, \fMAC$ signify the key size reductions during sifting, parameter estimation, privacy amplification and authentication, respectively. In the considered COW implementation, a fraction $\fSift=\left(1-\pDecoy\right)/\left(1+\pDecoy\right)$ of all detections in $\Ddata$ is discarded during sifting. Furthermore, it is $\fEst=0.875$ if we perform parameter estimation based on sub-sampling, and $\fEst=1$ if we estimate the QBER by key comparison. 

Including finite-key-size effects, the secret key fraction $\fSec$ under the assumption of a restricted collective attack \cite{Branciard2008} is given for a QBER $\QBER$ by the Devetak-Winter bound
\begin{align}
	&\fSec= 1-\text{leak}_{\text{EC}}-\text{leak}_{\text{VER}}\nonumber\\
	&-(\QBER+\deltaQ)-(1-\QBER-\deltaQ)\cdot h\left[\frac{1+\Delta }{2}\right]-\fSmooth-\fEC-\fPA
	\label{eq:fSecCOWFinite}
\end{align}
The leakage of the error correction scheme $\text{leak}_{\text{EC}}$ is in the ideal case the binary entropy $h\left[\QBER\right]$, while in the implementation at present, $\text{leak}_{\text{EC}}=1-f_{\text{EC}}$, with the chosen LDPC code rate $f_{\text{EC}}\in\left\{5/6, 3/4, 2/3, 1/2\right\}$. The leakage from the verification step after error correction amounts to $\text{leak}_{\text{VER}}=l/b=0.023$ with $l=48$~bits the length of each verification hash tag, and $b=2048$~bits the block length per verification. The overlap $\deltaLap =\left|\left\langle \psi_{\text{1}}|\psi_{\text{0}}\right\rangle\right|$ between the two bit states is for an observed visibility $\Vis$
\begin{align}
	\deltaLap =& (2\cdot\left(\Vis-\deltaV\right)-1)\cdot e^{-\muQ }\nonumber\\
	-&2\cdot \sqrt{1-e^{-2\cdot \muQ }}\cdot \sqrt{\left(\Vis-\deltaV\right)\cdot (1-\left(\Vis-\deltaV\right))}
\end{align}
Due to the finite post-processing size we include statistical fluctuations of expected QBER and visibility values, given by analysis based on interval estimation. For parameter estimation based on sub-sampling, it is~\cite{Sano10, Tomamichel12}
\begin{align}
	\deltaQ	&=\sqrt{\frac{1+\etaPE \cdot \left(\nPE-1\right)}{\left(\etaPE\cdot \nPE\right)^2}\cdot
\text{Log}\left[\frac{1}{\epsilon_{\text{PE}}}\right]}
	\label{eq:finDeltaQ}
\end{align}
In contrast, for parameter estimation based on key comparison, no uncertainty from statistical fluctuations impair the QBER, i.e. 
\begin{align}
	\deltaQ	&=0
\end{align}
However, in both cases the deducible visibility is limited by an uncertainty $\deltaV$ due to the finite-key-size as
\begin{align}
\deltaV	&=\sqrt{\frac{1}{2}\cdot\left(\text{Log}\left[\frac{1}{\epsilon _{\text{PE}}^V}\right]+2\cdot \text{Log}\left[n_{\text{V}}+1\right]\right)/n_{\text{V}}}
	\label{eq:finDeltaV}
\end{align}
$n_{\text{V}}$ is the number of useful detections in the monitor detector from which the visibility is calculated. In the trusted detector scenario the secret key rate is optimized using QBER and visibility values that are corrected for detector errors, which can not be exploited or manipulated by an eavesdropper, e.g. dark counts. For the leakage term in eq.~(\ref{eq:fSecCOWFinite}), the uncorrected QBER value must be considered. 

Furthermore, we account in equation~(\ref{eq:fSecCOWFinite}) for the reduction $\fSmooth$ due to uncertainty induced by smoothing the min-entropy, and the failure probabilities $\fEC$ and $\fPA$ of the error correction and privacy amplification protocols~\cite{Tomamichel12}
\begin{align}
	\fSmooth&=7\cdot \sqrt{\log_{2}\left[\frac{2}{\epsilonSmooth}\right]/\nPE}\label{eq:fSmooth}\\
	\fEC&=\text{Log}_2\left[\frac{2}{\epsilonEC}\right]/\nPE\label{eq:fEC}\\
	\fPA&=2\cdot\text{Log}_2\left[\frac{1}{\epsilonPA}\right]/\nPE\label{eq:fPA}
\end{align}
where the respective $\varepsilon$-parameters specify the confidence interval. For the presented implementation, the key length after parameter estimation $\nPE=\fEst\cdot\nSift$ equals the sifted key rate as no bit values are revealed for estimating $\QBER$. Instead, the errors are measured by comparing the original bit string with the corrected one, which limits $\epsilonEC$ to the confidence interval of subsequent error verification ($\epsilonEC=\epsilonVER=8\cdot 10^{-11}$). The total security parameter of the system is then fixed by the sum
\begin{align}
	\epsilonQKD=\epsilonSecr=\epsilonVisPE+\epsilonSmooth+\epsilonPA+2\cdot\epsilonVER+\epsilonMAC = 4\cdot 10^{-9}
\end{align}
Note the factor of two for $\epsilonVER$ to account for failures in the QBER measure as well as the verification step.

As a first input parameter we fix the number of bits $\nSift$ after sifting entering the further distillation post-processing which in our system is limited by the allocated hardware memory to $\nSift = 995,328$~bits. From this number of bits the respective number of useful detections $n_{\text{V}}$ in the monitoring detector which is used to estimate the visibility is derived as
\begin{align}
	n_{\text{V}}=\nSift\cdot \frac{\pDecoy+ \frac{\left(1+\pDecoy\right){}^2}{4} }{1-\pDecoy}\cdot
\frac{\left(1-\tBob\right)}{\tBob}
\end{align}
Here, the first factor is the normalization since we use all useful monitor detections, the second factor specifies the number of useful events due to decoy sequences and combinations across bit separations, and the third factor accounts for the beam splitting ratio. Any additional losses or differences in the detection efficiencies between data and monitor detector can be incorporated by a respective choice of the beam splitting ratio $\tBob$ and detection efficiency $\etaD$. Note that hypothetically, we assume an additional detector at the bright interferometer port, however, in practice this detector is not necessary.
%
\section*{References}
\bibliographystyle{unsrt}
\bibliography{Bibliography}
\end{document}